\documentclass[preprint,number,12pt]{elsarticle}

\usepackage{verbatim}
\usepackage{array}
\usepackage{cite}
\usepackage{url}
\usepackage{graphicx}
\usepackage{algpseudocode}
\usepackage[caption=true,font=footnotesize]{subfig}
\usepackage{amssymb}
\usepackage{amsmath}
\usepackage{setspace}
\usepackage{epstopdf}
\newsavebox{\ieeealgbox}
\newenvironment{boxedalgorithmic}
{\begin{lrbox}{\ieeealgbox}
\begin{minipage}{\dimexpr\columnwidth-2\fboxsep-2\fboxrule}
\begin{algorithmic}}
{\end{algorithmic}
\end{minipage}
\end{lrbox}\noindent\fbox{\usebox{\ieeealgbox}}}
\journal{Journal of Network and Computer Applications}

\begin{document}

\begin{frontmatter}



\title{Highly Available Smart Grid Control Centers through Intrusion Tolerance}


\author[a]{Maryam Tanha}
\author[a]{Fazirulhisyam Hashim\corref{cor1}}
\ead{fazirul@eng.upm.edu.my}
\author[b]{S.Shamala}
\author[a]{Khairulmizam Samsudin}
\address[a]{Faculty of Engineering, Universiti Putra Malaysia, Malaysia}
\address[b]{Faculty of Computer Science and Information Technology,  Universiti Putra Malaysia, Malaysia}
\cortext [cor1]{Corresponding author. Tel :  +603-8946 4319.}

\begin{abstract}
Societies' norms of operation relies on the proper and secure functioning of several critical infrastructures, particularly modern power grid which is also known as smart grid. Smart grid is interwoven with the information and communication technology infrastructure, and thus it is exposed to cyber security threats. Intrusion tolerance proves a promising security approach against malicious attacks and contributes to enhance the resilience and security of the key components of smart grid, mainly SCADA and control centers. Hence, an intrusion tolerant system architecture for smart grid control centers is proposed in this paper. The proposed architecture consists of several modules namely, replication \& diversity, compromised/faulty replica detector, reconfiguration, auditing and proxy. Some of distinctive features of the proposed ITS are diversity as well as the combined and fine-grained rejuvenation approach. The security of the proposed architecture is evaluated with regard to availability and mean time to security failure as performance measures. The analysis is conducted using a Discrete Time Semi Markov Model and the acquired results show improvements compared to two established intrusion tolerant architectures. The viability of SLA as another performance metric is also investigated.
\end{abstract}


\end{frontmatter}


\section{Introduction}
\label{introduction}
In recent decade, the growing dependence of critical infrastructures on Information and Communication Technology (ICT) and open standards has raised serious concerns about security issues. Future power grid also known as smart grid exemplifies one of these critical infrastructures. In addition to environmental benefits by using renewable energy resources to reduce the carbon footprint as well as the economic merits for both utilities and consumers (through dynamic pricing and active end-user participation), an outstanding feature of smart grid is the integration of fast, dependable and secure data communication networks to control and monitor the intricate power systems in an effective and intelligent way\citep{Wang20113604}. The cyber-physical dependencies (i.e., the combination of the legacy power grid and the communication networks and their interdependencies)\citep{Chen2012}, large-scale operation, heterogeneity and complexity\citep{Mo2012,Amin2005,Liu2012a} along with sophisticated and novel attacks pose grave and new threats to the mission critical applications in particular the smart grid. Moreover, the security objectives of the smart grid differ from the ICT security goals in their order of significance. Availability and continuity of service is the main security priority\citep{Liu2012}. Even the Quality of Service (QoS) requirements are different from ICT pre-requisites. Message delay is of great importance in smart grid whereas the data throughput receives special attention in the Internet\citep{Yan2012}. On top of all the mentioned issues, the widespread and socioeconomic impacts of malfunction or failure of the smart grid resulting from accidental or malicious events mandate more automatic and robust security solutions\citep{Verissimo2006,Bessani2008c}.

The prime goal of any cyber-physical system such as smart grid is to offer smooth control over some physical process \citep{Sridhar2012} which will result in considering availability and integrity as the overriding security attributes. Thus, attacks on the control systems of cyber-physical systems can  adversely affect their security and reliability. Some of the recent high-profile attacks have been mainly targeted at critical infrastructure control systems and crucial organizations. The Stuxnet worm, emerged in July 2010, aimed to control critical infrastructures. It exploited a vulnerability in MS Windows and attempted to modify the code running in Programmable Logic Controllers (PLCs)\citep{Karnouskos2011,Seo2011}. Duqu, nearly a similar malware to Stuxnet, came into the spotlight in 2011. It acted as a Remote Access Trojan (RAT) with the purpose of information gathering but not specifically targeted at control systems. Nevertheless, among the compromised organizations were “manufacturing of industrial control systems”\citep{tagkey20113,tagkey20123}. The most recent and novel threat (made public in May 2012), called FLAME\citep{SKyWIperAnalysisTeam2012}, was a sophisticated and state-of-the-art malware with a modular structure able to perform information theft and took advantage of many attack and propagation methods. This espionage malware's end was some governmental organizations in Iran\citep{Xxa}.

All the aforementioned security concerns and incidents serve as contributing factors to change our mind set about the level of security that can be achieved through present security mechanisms especially for critical infrastructures. Conventional security approaches (i.e., prevention and detection) proved insufficient to tackle the security problems\citep{Bessani2008c}, thus underscoring the need for a more resilient and robust security approach. To satisfy the mentioned security requirements, a promising mechanism called intrusion tolerance has come to existence. Thus, this approach to security has received considerable attention in recent years\citep{Verissimo2006,Bessani2008c,Nguyen201124,Raj2011,Uemura2010,Bessani2011a,Sousa2010,Deswarte2006432,Nagarajan2010a,Bangalore2009a,Saidane2009a}. However, as stated in\citep{springerlink:10.1007/3-540-45177-3_1}, the first usage of the term “intrusion tolerance” dates back to 1985 in\citep{Fraga1985203} by Fraga and Powell. Intrusion tolerance is concerned with the fact that there is always probable for a system to be vulnerable to security compromise as well as for some attacks to be launched successfully on a system\citep{Verissimo2006}. In spite of these assumptions, intrusion tolerance mechanisms ensure that the system prolongs its normal activities (or acts in a degraded mode providing only essential services) even when it is under attack or partially compromised. Thus, rather than preventing intrusions from happening in the system, they are permitted but tolerated by adopting and triggering appropriate mechanisms such as redundancy, diversity, rejuvenation, and so on. These techniques result in masking, removing or recovering from intrusions and preclude them from turning into security failures\citep{springerlink:10.1007/3-540-45177-3_1}. Consequently, the system remains highly survivable to malicious attacks and intrusions. Therefore, intrusion tolerance can be considered as a last resort security solution when other security measures fail to accomplish their intended purpose.

 To the best of our knowledge the  significance of intrusion tolerance as a prospective security mechanism for smart grid has only been pointed out in the research carried out in\citep{Sridhar2012} and\citep{Overman2011c}. This paper highlights the importance of intrusion tolerance approach which raises the possibilities for enhancing the security of critical components in the smart grid, particularly control centers and Supervisory Control and Data Acquisition (SCADA) systems. Hence, an Intrusion Tolerant System (ITS) architecture is proposed to strengthen and enhance the level of security in such systems. In addition, the security attributes of the proposed architecture are evaluated using a semi Markov model.

The paper is organized as follows. Section~\ref{sec: smart grid cyber security} highlights the security of the smart grid communication infrastructure especially in control centers. This section places an emphasis on the need for a robust defense-in-depth security approach to be adopted in smart grid control centers taking into account the limitations of the fundamental security mechanisms. Therefore, intrusion tolerance is introduced as a promising security solution for smart grid. Section~\ref{sec: intrusion tolerance for smart grid} provides a detailed analysis of intrusion tolerance. The difference between intrusion tolerance and fault tolerance along with classical security mechanisms (i.e., prevention and detection) are deliberated. The most commonly used intrusion intrusion tolerance techniques are presented as well as a comparison is made between some of existing ITS architectures. In Section~\ref{sec: proposed architecture}, a detailed discussion on the proposed intrusion tolerant architecture for smart grid control centers is presented. The performance of the proposed architecture is evaluated analytically and compared with established ITSs in Section~\ref{sec: performance analysis}. Finally, Section~\ref{sec: conclusion} draws the conclusion.
\section{Smart Grid Cyber Security as a Cyber Physical System}
\label{sec: smart grid cyber security}
\begin{figure}[!t]
\centering
\includegraphics[width=3.4in]{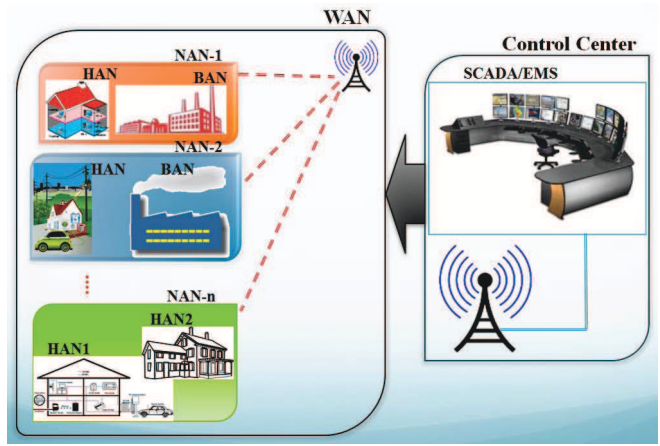}
\caption{Smart grid communication infrastructure.}
\label{fig:SmartGridInfra}
\end{figure}

As we mentioned in the earlier section, smart grid is the modernized power grid that is inextricably interwoven with information and communication technology. Advanced features of such a complex system of systems include, but not limited to, two-way communications between customers and utilities, Demand Response (DR), Distributed Energy Resources (DER), sophisticated sensing technologies and real-time control and monitoring\citep{Fan2012} along with self-healing capabilities. Enabling these functionalities requires an effective, reliable, secure and resilient communication infrastructure\citep{Li2012}. Figure~\ref{fig:SmartGridInfra} demonstrates a general view of smart grid communication infrastructure. Home Area Networks (HANs) and Business Area Networks (BANs) comprise the level of communication infrastructure which is in close proximity to the electricity consumers. This section of the smart grid communication infrastructure enables DR and active participation of end users through the use of smart meters. Geographically close HANs or BANs make up another level of the smart grid infrastructure hierarchy called neighborhood area network (NAN). It contributes to the exchange and sharing of information between electricity distribution facilities and consumers' premises. Finally, Wide Area Network (WAN) furnishes the smart grid infrastructure with the backbone to transmit control commands and monitoring signals from control centers, SCADA systems in particular, to electric devices located in substations as well as the real-time measurements from electric devices to the control centers. It encompasses several NANs each formed under one substation\citep{Wang20113604}.
\captionsetup[figure]{labelsep=period}
\captionsetup[table]{labelsep=period}

Smart grid is regarded as a cyber-physical system, i.e., it is a system in which cyber security attacks can give rise to disruptions that go beyond the cyber domain and impact the physical world\citep{Mo2012}. In other words, security attacks may occur in both the physical space (i.e., the traditional power grid) and cyber space (i.e., the communication networks). Security in both physical and cyber domains is one of the principal objectives of the smart grid\citep{Li2012}. The U.S Department of Energy (DoE) has recognized attack resistance as one of the salient features needed for the operation of the smart grid\citep{Chen2012}. Using open standard software and protocols have opened avenues for attackers to pose dire threats to different sections of smart grid communication infrastructure particularly, SCADA systems. Furthermore, the escalating number of electrical outages and brown-outs worldwide during the last decade proved the power grid to be a potential target for malicious attacks. The cascaded power outages have arisen in Europe\citep{Pearson20115211} as well as the ones come up in the United States and other countries\citep{Amin2005,Amin2012} are the consequences of such intrusive measures.

\begin{figure}[!t]
\centering
\includegraphics[width=3.4in]{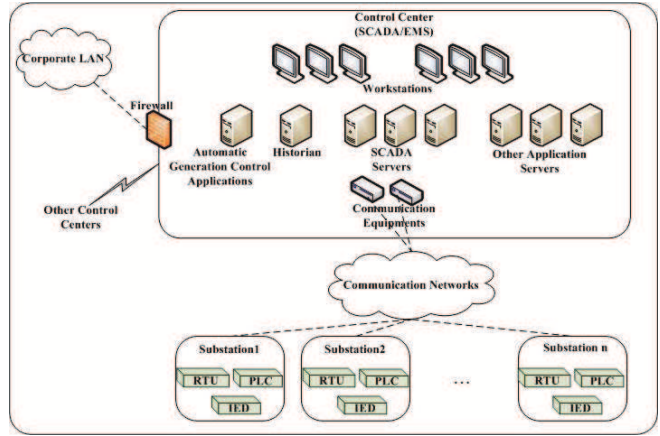}
\caption{Control center and substations in smart grid.}
\label{fig:ControlCenter}
\end{figure}

Control centers are considered as the brain of the smart grid\citep{Zeng2012}. They are in charge of data analysis and decision making\citep{Chen2012}. Based on the assembled data, they make appropriate adjustments to power supply to satisfy demand as well as spot and respond to the defects or failures by sending control commands to field devices\citep{Li2012}. SCADA and Energy Management System (EMS) as the key components in control centers play a pivotal role in the proper operation of smart grid, any malfunction or failure of these systems may result in widespread and devastating effects (e.g., power outage, cascading blackouts) on industry, economy and people's daily life. Other possible consequence of the control centers disruption is loss of consumer and public trust\citep{Zeng2012}. Therefore, the correct functioning of these systems in exigent security circumstances are of paramount importance. Several attacks on SCADA systems have been investigated in\citep{Nicholson2012418} according to their perpetrators as well as the industry sectors influenced by such attacks. Energy sector had been noticeably impacted by the SCADA incidents and attacks compared to the other industry sectors. The risks associated with the SCADA system that a government or company may faced with can be considered as financial loss or even injury or loss of life. Defects such as unpatched software, software bugs, buffer overflows and poor administration contributes to launching attacks against  the SCADA systems. Two of the dire threats to SCADA/EMS are considered as Denial of Service (DoS) and unauthorized access/integrity breach\citep{Nicholson2012418}. These threats will result in the unreliability of the control signals from the monitoring system in addition to the measurement data gathered in the smart grid that are used for pricing or state estimation purposes. The possible ramifications would be massive brownouts or blackouts. For instance, SQLSlammer worm  is a serious DoS attack against the control systems in the smart grid and any critical infrastructure. Thanks to the time-criticality of the communication and control in smart grid, a delay of a few seconds (following from an availability attack) may lead to irreparable harm to the national economy and security\citep{Li2012}. Figure~\ref{fig:ControlCenter} shows a control center which supervises multiple substations in smart grid\citep{Wei2011c}. The key components of the control center (i.e., SCADA/EMS) and the substations, including PLC, Remote Terminal Unit (RTU) and Intelligent Electronic Device (IED) are shown as well.

The significance of a defense-in-depth security approach for smart grid has been highlighted in several published papers\citep{Seo2011,Overman2011c,Amin2012,Ericsson2010} since it requires the adversaries to spend a great deal of time and effort to evade different layers of defense. This layered approach would involve the adoption of best cyber security practices such as firewalls, Role-Based Access Control (RBAC), cryptography, Intrusion Detection and Prevention Systems (IDPS), and so on. However, these security mechanisms are subject to certain restrictions concerning their scope of operation and effectiveness. In addition to the limitations associated with the classical security approaches, there are other factors that impose restrictions on using some of these mechanisms in SCADA systems. Since the SCADA systems hinge on timely presentation of data, firewalls and anti-viruses may reduce the speed of data flow and subsequently lead to decreasing the accuracy of SCADA systems. In such circumstances, the SCADA operators tend to deactivate or bypass these security mechanisms. Furthermore, patching SCADA systems is a tricky task due to introducing unknown impacts into the system that probably violate their correct operation and availability as well as the lack of comprehensive test environments. Network Intrusion Detection Systems (NIDSs) may compensate for such a limitation\citep{Nicholson2012418}.
\section {Cohesive Intrusion Tolerance for Smart Grid}
\label{sec: intrusion tolerance for smart grid}
As stated in the previous section, intrusion tolerance shows enormous potential to be adopted and deployed in smart grid control centers. Intrusion tolerance and its paradigms enable secure and normal operation of the smart grid control centers, even when the system is being attacked or partially compromised. The primary goal is to tolerate malicious events and sustained attacks as well as masking, removing or recovering from intrusions. Thus, intrusion tolerance measures avert security failures and aid to maintain the availability of the system. Moreover, intrusion tolerance places emphasis on the impact of the attack rather than the cause of it.
\subsection{A General Misconception about Intrusion Tolerance}
Some people may have a preconception about intrusion tolerance that leads to considering fault tolerance and intrusion tolerance as the same concepts. But in fact, fault tolerance can be considered as a predecessor of intrusion tolerance. Although these mechanisms have similarities especially in the techniques that they are using, some differences exist. The distinction between these two approaches lies in the nature of possible faults in a system. Fault is an imperfection or defect in the system that can give rise to an error which may result in a subsequent failure. Based on the definition of fault in\citep{Deswarte2006432,Sterbenz20101245},
 it is viable to categorize faults into non-malicious and malicious. Non-malicious faults include accidental design flaws (e.g., software bug), deliberate design defects following from constraints such as cost, environmental or natural perturbations or even a mistakenly action carried out by a distracted operator. It is apparent that these types of faults are infrequent and occur at random. Fault tolerance deals with this class of faults. In contrast, a malicious fault, also called an intrusion, is an intentional operational fault that stems from a successful attack on a system vulnerability. Intrusion tolerance should handle malicious faults, i.e., intrusions that are prevalent in information and communication systems and also critical infrastructures such as smart grid.
\subsection{Intrusion Tolerance versus Classical Security Mechanisms}
Intrusion tolerance is commonly referred to as the third generation of security technologies\citep{springerlink:10.1007/978-3-642-23971-7_36} which provides complementary features to conventional security mechanisms, i.e., prevention and detection. Some of the driving forces behind the increasing tendency for employing intrusion tolerance techniques are as follows:
\begin{itemize}
\item The growing number of novel and zero-day attacks and, thus the infeasibility to prevent or detect all intrusions in an effective manner\citep{Verissimo2006}
\item The sheer complexity of the systems that makes it impossible to pinpoint all of their vulnerabilities prior to coming into operation \citep{Saidane2009a}
\item Preventive security measures such as firewall, access control, authentication and authorization mechanisms are mainly proactive \citep{Uemura2010} and do not guarantee perfect protection\citep{Verissimo2006}
\item Intrusion detection techniques including misuse detection  and anomaly 
detection are based on property checks (i.e., comparing observed activity with 
known patterns of attacks or normal behavior of the system)
\citep{Deswarte2006432} and may result in high false positive or false negative 
rates
\item  Detection methods are predominantly reactive with limited automated 
defense capabilities and require human intervention to conduct a post-mortem 
and deal with the identified security threats\citep{Wang2003b}. This may lead 
to a slow reaction in the face of attacks that require to be dealt with immediately (especially in critical systems such as smart grid control centers)
\end{itemize}

In regard to the aforementioned issues and the fact that downtime, failure or malfunction of the smart grid control centers is not acceptable and must be kept at minimum, there is an urgent need for more automatic and resilient security approaches. Therefore, intrusion tolerance through appropriate means (e.g., redundancy, dynamic and adaptive rejuvenation and reconfiguration) is vital to fulfill the security and survivability requirements of the smart grid.
\subsection{Paradigms of Intrusion Tolerance}
Figure~\ref{fig:IntrusionToleranceParadigms} illustrates several common paradigms of intrusion. These techniques assist in achieving the goal of intrusion tolerance which is provisioning correct service despite the presence of active attacks and intrusions. Although utilizing these methods incurs substantial costs such as performance costs, administration and maintenance costs, no expense is spared employing them in mission critical systems such as smart grid control centers in which adverse effects of intrusions may lead to higher expenses or even irrecoverable losses. The description of the most widely used paradigms of intrusion tolerance are as follows:
\begin{figure}[!t]
\centering
\includegraphics[width=3.4in]{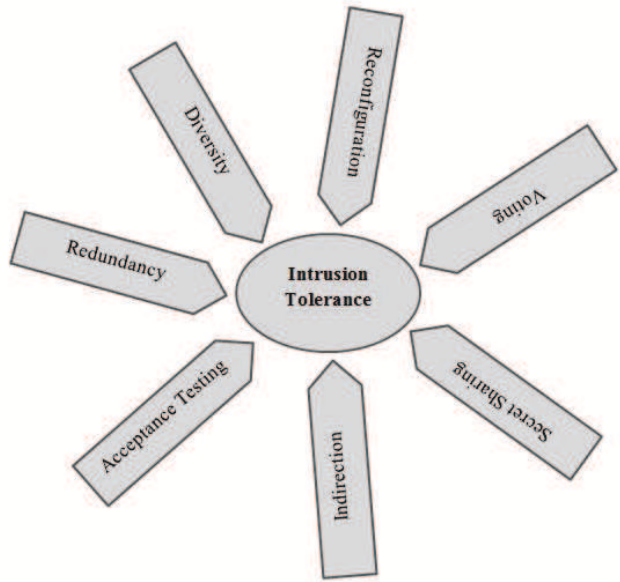}
\caption{Paradigms of intrusion tolerance.}
\label{fig:IntrusionToleranceParadigms}
\end{figure}
\subsubsection{Redundancy}
Redundancy is defined as allotting additional resources to a system that are more than its usual needs in normal functioning situations. There exists different types of redundancy including space redundancy, time redundancy and information redundancy among which space redundancy (i.e., physical resource redundancy or replication) has received considerable attention. In fact, replication is an indispensable part of intrusion tolerant systems. Replicated systems usually operate based on Byzantine Fault Tolerant (BFT) protocols, i.e., a system contains $n$ replicas is able to work properly even if $f<n$ replicas undergo arbitrary faults ($f$ is the number of tolerable faulty replicas)\citep{Bessani2011a}. However, redundancy suffers from the underlying problem of fate sharing for replicas\citep{Sterbenz20101245,Wang2003b}. If an adversary locates and exploits a vulnerability in one replica, it is highly likely that all replicas are susceptible to the same threat.
\subsubsection{Diversity}
To alleviate the problem of fate sharing associated with redundancy, it is common to employ diversity as a complementary technique to redundancy. Diversity equips the replicas with security failure independence. Diversity also has some variants such as space diversity, time diversity and implementation diversity. Operating System diversity (OS diversity) has gained momentum to be adopted in intrusion tolerant systems. The reason is twofold. First, the availability, less complexity and cost-effectiveness of using various operating systems to provide diversity makes them an appropriate choice for applying diversity to ITSs. Second, many intrusions are concerned with vulnerabilities of different operating systems due their pivotal role in any system\citep{Garcia2011a}. Operating system can be considered as one of the vulnerable parts of a system regardless of the robustness of the software running on top of it\citep{Obelheiro2006}.
\subsubsection{Dynamic recovery and reconfiguration}
One of the contributing factors to intrusion tolerance is to dynamically reconfigure the replicas. This reconfiguration may involve measures such as rejuvenation (i.e., recovery), modifying the system's posture, rollover and load sharing among which rejuvenation is widely employed in ITSs. Rejuvenation involves the restoration of a replica to a pristine state to eliminate the likely effects of intrusions or faults\citep{Wang2003b}. This may include the modification of the cryptographic keys or loading a clean copy of the operating system and applications\citep{Sousa2010}.

Although the BFT protocols are effective in holding up the failure of the overall replicated system by a specific amount of time, they are highly dependent on the value of $f$ and the degree of diversity in the replicated system\citep{Bessani2011a}. Increasing the $f$ incurs more cost on the system as well as it necessitates the rise of diversity degree which has a limited scope (e.g., in case of OS diversity, the number of existing operating systems  are limited). Rejuvenation is considered as an acceptable solution to the mentioned problems by decreasing the value of $f$ and the duration of time the attacker has at his disposal to corrupt more than $f$ replicas. Moreover, the constant availability requirement along with the unknown execution time of critical infrastructures such as smart grid underline the need for a kind of recovery mechanism to make sure that the allowed maximum number of compromised components is not violated. However, for the rejuvenation to be effective, the rejuvenation rate should be kept more than the intrusion rate\citep{Sousa2006}. Another point is that the allowed number of faulty or compromised replicas (i.e., $f$) is an upper bound on the number of concurrent rejuvenations. The availability is violated when the total number of compromised replicas and the rejuvenations in progress exceeds $f$. Therefore, the total number of replicas should be more than $n$ (as in BFT systems) to satisfy the availability requirements\citep{Sousa2008}.

Rejuvenation would have a superior impact on enhancing the security of the system if it is combined with diversity (e.g., restoring the replica with a clean version of a different OS). This is due to the fact that the recovery process may eliminate the impacts of fault or intrusion but there is no guarantee that the rejuvenated replica compromised again exploiting the same vulnerability. The situation would get aggravated if the adversary has gained critical information (e.g., passwords, OS version) prior to the rejuvenation that may result in a more sophisticated form of attack following the recovery.
Rejuvenation can possess two different forms as follows:
\begin{enumerate}
\item Proactive rejuvenation: Proactive recovery is the process of periodically rejuvenating the replicas. It assists in the identification of dormant faults (these faults may not even detected through detection mechanisms) or masking intrusions and should be conducted frequently sufficient to restrain the adversary from infecting more than $f$ replicas during a proactive recovery period (assuming that no reactive recovery performed in this period). The downside of this method is that it may not be effective in an asynchronous system since the attacker can postpone the recovery of a compromised replica by manipulating the system's clock. As a result, he/she will have adequate time to compromise more replicas than the system is able to tolerate\citep{Sousa2010}. Another possibility is that the attacker may be able to intrude the components at a rate faster than rejuvenation\citep{Sousa2006}.
\item Reactive rejuvenation: This kind of rejuvenation complements the 
proactive recovery by speeding up the process of handling compromised
 replicas. It is usually triggered by intrusion detection mechanisms to 
rejuvenate  the suspected and faulty replicas. If a compromised replica 
cannot be identified by the adopted detection mechanisms in the system, 
there is no way to signal the reactive recovery to be performed. Hence, 
the intrusion will go undetected without raising any suspicion \citep{springerlink:10.1007/978-3-642-23971-7_36}.
\end{enumerate}
\subsubsection{Voting}
Voting algorithms are employed to reach a consensus on the valid and final output of non-faulty redundant components in an ITS. Using some criteria such as edit distance (e.g., hamming distance) and hash codes make the comparison feasible. Voting contributes to masking and tolerating intrusions. Formalized majority voting and formalized plurality voting are some of these algorithms\citep{Wang2003b}.
\subsubsection{Secret sharing scheme}
Secret sharing or threshold scheme is based on the idea of concealing a piece of information by splitting it into several shares and distributing among participants in a manner that  specific subsets of the shares are required to rebuild the initial data\citep{AlEbri2011}. In regard to application in ITSs, the secret data can be the main information or the associated cryptographic key. The former entails storing the data shares in separate physical locations in a way that the confidentiality is maintained and the original information can be rebuilt even if a certain number of shares infected or compromised by attackers. In the second case, the key used to encrypt the data is broken down into shares so that a particular number of shares are needed to reconstruct the original key and access the data\citep{Wang2003b}.
\subsubsection{Acceptance testing}
Having different forms, including requirement test, reasonableness test, timing test, accounting test and coding test, acceptance testing is a programmer or developer-provided error detection function in a software module to inspect the reasonableness of the generated results. This technique similar to redundancy and diversity has its root in the fault tolerance. However, being application dependent is regarded as one of the drawbacks of this technique. Therefore, creating appropriate and effective tests necessitates understanding the system painstakingly. More details on the acceptance testing measures can be found in\citep{Wang20031399}.
\subsubsection{Indirection}
Proxies, wrappers and virtualization are some of the indirection techniques that serve as additional layers of defense between servers and clients. In spite of their benefits, they incur the cost of overhead and latency so these factors should be taken into account when designing the system\citep{Wang2003b}.

\begin{table}[!t]
\renewcommand{\arraystretch}{1.3}
\caption{Proposed comparative analysis of intrusion tolerant architectures. (Y: Yes, N: No, O: Optional)}
\label{table_ITS architecture comparison}
\centering
\scalebox{0.5}{
\begin{tabular}{c||c||c||c||c||c||c||c||c||c}
\hline
\bfseries & \bfseries COCA & \bfseries DIT & \bfseries Willow & \bfseries SITAR & \bfseries SCIT & \bfseries MAFTIA & \bfseries Crutial & \bfseries FOREVER &\parbox{2cm}{\bfseries Generic\,ITS\\ \, for\,web\\ \,servers}\\
\hline\hline
Replication & Y & Y & Y & Y & Y & Y & Y& Y& Y\\
\hline\hline
Diversity & N & Y & Y & Y & O & Y & Y& Y& Y\\
\hline\hline
Proactive Recovery & Y & Y & Y & N & Y & N & Y& Y& Y \\
\hline\hline
Reactive Recovery& N & Y & Y & Y & N & Y & Y& Y& Y\\
\hline\hline
Fine-grained Recovery &N & N & N & N &N & N & N&N&N\\
\hline\hline
Voting/BFT Agreement & Y & Y & N& Y & N &Y & Y& Y&N\\
\hline\hline
Proxy&N & Y & N & Y & N & N& N& N& Y\\
\hline\hline
Intrusion Detection Capabilities & Y & Y & Y & Y & N & Y & Y& Y& Y\\
\hline\hline
Secret Sharing & Y &N & N & N & N & Y & N&N& N\\
\hline
\end{tabular}}
\end{table}

\subsection{Intrusion Tolerant Architectures}
During the last decade, various research have been conducted on intrusion tolerance and multiple intrusion tolerant architectures with specific features and applications have been proposed. The Willow architecture\citep{Knight2003a}, COCA\citep{Zhou2002}, DIT\citep{Valdes2003}, MAFTIA\citep{Stroud2004a} , SITAR\citep{Wang2003a}, SCIT\citep{Bangalore2009a}, Crutial\citep{Bessani2008c}, FOREVER \citep{Sousa2008} and Generic intrusion tolerant architecture for web servers\citep{Saidane2009a} exemplify a number of the proposed ITS architectures. Some of these architectures are application-specific. For instance, the goal of COCA is to provide a secure and fault-tolerant certification authority (CA) while Crutial is a distributed firewall-like intrusion tolerant system for critical infrastructures protection such as power grid. But primarily, enhancing the security and availability of distributed services, Commercial Off The Shelf (COTS) servers and critical information systems have called for designing such architectures. The intrusion tolerance paradigms introduced in the previous section are commonly used in intrusion tolerant systems. Hence, they can be utilized to analyze and compare different intrusion tolerant architectures. Some representatives of existing ITS architectures have been compared by conducting qualitative analyses in\citep{Nguyen201124,Raj2011}. Table \ref{table_ITS architecture comparison} depicts such analysis but with emphasis on the paradigms of intrusion tolerance employed in several ITSs. The spectrum of architectures have distinct features. We have developed a comparative analysis to enable a clear reflection of their respective attributes. Moreover, our provided comparison encompasses a higher volume of ITSs.  As it can be seen, replication and diversity are the techniques adopted by almost all the ITSs. Although design diversity (e.g., using different versions of OS) is the dominant type of diversity used by the ITSs, FOREVER and Crutial can employ time diversity (i.e., rejuvenation introduces diversity). Some ITSs such as FOREVER, Willow and Crutial apply a combined recovery method, that is, both proactive and reactive recovery whereas others like Self Cleansing Intrusion Tolerance (SCIT) only use proactive recovery. To the best of our knowledge, none of the existing ITS utilizes the fine-grained recovery strategy introduced in\citep{springerlink:10.1007/978-3-642-23971-7_36}. One of the indirection techniques that is widely preferred is the use of proxies as the mediator between the COTS servers and the outside network. Intrusion detection whether anomaly-based or signature-based are very common among the ITSs. Byzantine agreement algorithms and secret sharing are the other intrusion tolerant mechanisms that have been implemented in some of the architectures. Among the ITSs shown in Table 1, SCIT and Scalable Intrusion Tolerant Architecture for Distributed Services (SITAR) have drawn more attention in published intrusion tolerance research and investigated with regard to their performance\citep{Garcia2011a,Obelheiro2006,Sousa2006,Sousa2008}.

One of the formidable challenges that the ITSs must handle is their self-security against malicious manipulations and attacks. Therefore, several methods/modules in the ITSs are employed to meet this challenge some of them are as follows:
\begin{itemize}
\item Sensor subsystem, runtime verification and private subnet between the proxy and other components in DIT architecture
\item Audit control in SITAR
\item One-way signal from controller to the servers in SCIT
\item Distributed trust throughout the system in MAFTIA
\item Wormhole in Crutial and FOREVER
\item Runtime verification in the Generic ITS for web servers
\end{itemize}

Other important issues that should be addressed include the complexity, performance and cost. For instance, the relative complexity of SITAR is high whereas the associated complexity of SCIT is low\citep{Nguyen201124}. Desired performance metrics are chosen with regard to the application. In case of the smart grid, increased complexity of the ITS may be an advantage for the system since it makes it difficult for the attackers to break into the system. However, the complexity of a proposed ITS architecture for smart grid should not considerably degrade the desired performance of it. For instance, delay is of paramount importance for the communications from control centers to substations\citep{Wei2011c}. In addition, the degree of redundancy and diversity are policy-dependent and should be set at the deployment time.
\section{Proposed ITS Architecture for Smart Grid Control Centers}
\label{sec: proposed architecture}
\begin{figure}[!t]
\centering
\includegraphics[width=3.4in]{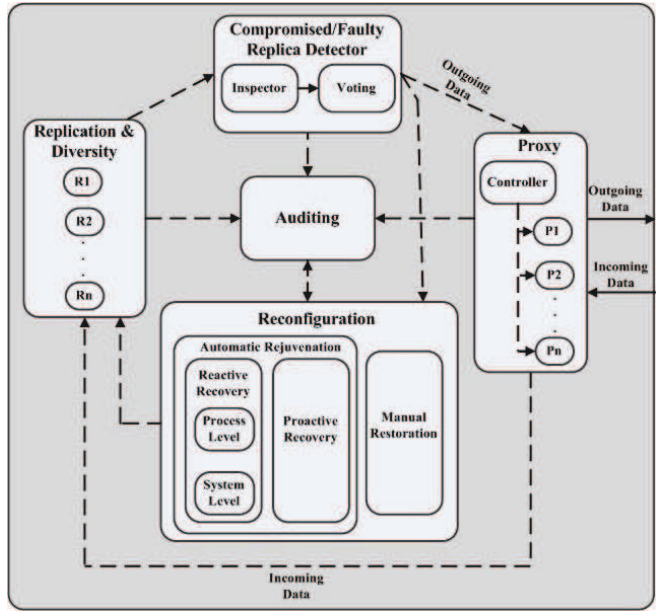}
\caption{The proposed ITS Architecture.}
\label{fig:proposedITS}
\end{figure}

Typical ITSs have single primary focus such as SITAR, which is detection triggered, and SCIT, which is recovery based. Moreover, as it has been shown in Table 1, none of the existing ITS architectures employs the fine-grained rejuvenation approach. These issues along with the specific requirements of the smart grid control centers (e.g., delay sensitivity) underscore the need for a new ITS architecture that suits the smart grid control centers. The proposed ITS for SCADA systems in smart grid encompasses a rich blend of a wide spectrum of different intrusion tolerance techniques. As illustrated in Figure~\ref{fig:proposedITS}, the proposed system comprises five modules, namely replication \& diversity module, auditing module, compromised/faulty replica detector module, reconfiguration module and proxy module. The role and working principles of the aforementioned modules are elucidated in the following sections. It should be noted that to avoid the proposed ITS from being compromised by the intruders, it is assumed that all the components' tasks and their communications are performed in a trusted platform. Proxy module also helps to enhance the security of the ITS.
\subsection{Replication and Diversity Module}
The replication module consists of a number of replicas for a critical entity in the SCADA systems of the smart grid such as Master Terminal Unit (MTU). The number of replicas should be at least $2f + 1$ to tolerate $f$ intrusions. In this module, the number of replicas is assumed to be $2f + 1 + K$ and the value of $f$ ($f\geqslant1$) and $k$ are indicated in the deployment time. The same approach also used to design a distributed firewall-like protection device named Crutial Information Switch (CIS) in\citep{Bessani2008c}. The reason why the value of $k$ is added to the number of replicas will be given in the reconfiguration module section. It should be noted that all replicas have OS diversity to decrease the probability of sharing the same vulnerabilities. OS is considered a vital element of each replica on account of hosting the SCADA system. Thus, any misconfiguration or vulnerability in OS may bring down the SCADA system and causes the adversaries achieve breakthroughs\citep{Nicholson2012418}.

\subsection{Compromised/Faulty Replica Detector}
This module aims at examining the responses/outputs of the replicas to identify possible infected/compromised ones. It is composed of the following two sub modules:
\begin{enumerate}
\item Inspector: Acceptance testing as an intrusion tolerance technique is entailed
in the inspector module. It involves application-specific checks with regard to the security policy to ensure the sanity of outgoing data from the replicas. Any symptom of security compromise  detected by it will trigger the reactive recovery sub module in the reconfiguration module.
\item Voting: This sub module is intended for masking the impacts of intrusions as well as ensuring the integrity of replicas’ outputs. Based on a voting algorithm, it seeks for the correct output by comparing the redundant outputs from the active replicas that passed the inspector successfully. In this way, it will arrive at a consensus on the final desired output to be passed to the proxy module. This output can be a command or information from the SCADA critical components destined for a particular field device in smart grid. The invalid outputs will trigger the reactive recovery sub module of the reconfiguration module for their corresponding replicas.
\end{enumerate}

\subsection{Reconfiguration Module}
Reconfiguration module consists of two sub modules namely, automatic rejuvenation and manual restoration. When the proposed ITS is able to mask an intrusion, it leverages the automatic rejuvenation sub module, otherwise it takes advantage of manual restoration which involves human intervention.
Manual restoration happens when for instance the system is targeted by DoS attacks and only capable of provisioning the essential services
(graceful degradation). The sub modules descriptions are as follows:
\begin {enumerate}
\item Automatic rejuvenation: In this module, a combined rejuvenation approach (i.e., proactive-reactive recovery) has been used to compensate for the shortcomings of the two aforementioned rejuvenation approaches, i.e, reactive and proactive recovery. Thus, it will enhance the performance of the system through decreasing the possible duration of time a compromised replica may disrupt the normal operation of the system\citep{Sousa2010}. Automatic rejuvenation module enables the concurrent rejuvenation of at most $k$ replicas out of $2f+1+k$ (total number of replicas). The assumption for the total number of replicas eliminates the impact of compromised replicas (at most $f$) and recovery on the availability of the system. Proactive recovery (at the system level) is performed periodically through choosing an active replica based on smallest rejuvenation time stamp. Note that at most one replica is allowed to undergo this type of recovery at a time. Figure~\ref{fig:proactiveRecovery} shows the proactive rejuvenation mechanism. Reactive rejuvenation complements the proactive recovery. For reactive recovery, we have been inspired by a hierarchical recovery method that has been proposed recently in\citep{springerlink:10.1007/978-3-642-23971-7_36}. It encompasses three levels of recovery granularity, namely system level, object level and process level recovery. This model eliminates the need for complete recovery when the system is partly compromised. The merits of this model can be considered as reduced total recovery time, improved flexibility and dependability. In our system, reactive recovery can be triggered externally and at the system level by the compromised/faulty replica detector module (and may introduce diversity) or internally (within a replica) in a hierarchical and fine-grained fashion (including process level recovery and system level recovery) almost similar to the strategy proposed in\citep{springerlink:10.1007/978-3-642-23971-7_36}. The maximum potential number of replicas that can be under system level reactive recovery is $k$. Figure~\ref{fig:processLevelrecovery} depicts process level recovery. Process manager (can act as a type of Host-based IDS which features self-healing capabilities) is a module executed in each active replica to handle the process level recovery. There are two sets of processes, namely active set (includes running processes) and standby set. Based on a timeout period, the process manager examines the pool of active processes.  In the event of finding any suspected process, it will obtain the relevant checkpoint, kills the process and activates its peer from the standby set (if there is any) otherwise the system level recovery will be performed. The process level recovery is time-saving compared to system level recovery as well as it is more secure since it involves internal information and communication exchange in a machine. Moreover, it does not require the replica to go offline for performing the recovery.
\item Manual restoration: This sub module is triggered when the intrusion (whether detected or not) is non-maskable (e.g., more than $f$ replicas have been compromised). This may cause the system to be in graceful degradation mode, stopped functioning mode or complete failure mode all of which require human intervention and corrective measures to return the to the normal working state.
\end{enumerate}
\begin{figure}
\begin{boxedalgorithmic}
\Procedure{proactive-rejuvenation}{ }
\State  $Wait(RejuvenationPeriod);$
\While{$NoConcurrentRejuvenations\geq k$}
\State $Wait(RejuvenationPeriod);$
\EndWhile
\State $ i\gets Find(ReplicaWithSmallestTimestamp);$
\State $Replica[i].status\gets recovery;$
\State $NoConcurrentRejuvenations++;$
\State $Replica[i].SystemLevelRecovery();$
\State $Replica[i].SetTimestamp();$
\State $Replica[i].status\gets active;$
\State $NoConcurrentRejuvenations--;$
\EndProcedure
\end{boxedalgorithmic}
\caption{Proactive recovery mechanism.}\label{fig:proactiveRecovery}
\end{figure}
\begin{figure}
\begin{boxedalgorithmic}
\Procedure{process-manager}{ }
\While{$not\, timeout$}
\State $Wait ();$
\EndWhile
\State $Polling();$
\ForAll{the suspected processes (j) in replica i}
\If{$Process[j].StandbyAvailable()$}
\State $Process[j].ObtainCheckpoint(Suspect);$
\State $Process[j].Kill(Suspect);$
\State $Process[j].ActivateStandby();$
\Else
\If{$NoConcurrentRejuvenations<k$}
\State $Replica[i].status\gets recovery;$
\State $NoConcurrentRejuvenations++;$
\State $Replica[i].SystemLevelRecovery();$
\State $Replica[i].SetTimestamp();$
\State $Replica[i].status\gets active;$
\State $NoConcurrentRejuvenations--;$
\State $Exit\, the\, for\, loop$
\EndIf
\EndIf
\EndFor
\State $timeout = False;$
\EndProcedure
\end{boxedalgorithmic}
\caption{Process level recovery in a replica.}\label{fig:processLevelrecovery}
\end{figure}

\subsection{Auditing Module}
This module maintains audit logs for all modules. The logs would be useful for security administrator to monitor and analyze the operation of the system. 
\subsection{Proxy Module}
The proxy module is placed on the boundary of the ITS architecture where the data comes in or goes out of the intrusion tolerant architecture. The proxy module shields the internal structure of the ITS from attackers as well as acting as a load balancer. When the state information of field devices or power usage data collected by smart meters (incoming data in Figure~\ref{fig:proposedITS}) gathered in field devices passed to the respective critical components in the control center, they go through the proxy module as the first layer of defense. This incoming data is then forwarded to the replication and diversity module to be dealt with. Moreover, the control commands from the SCADA system (outgoing data in Figure~\ref{fig:proposedITS}) pass the proxy to reach the field devices in substations. Proxy module is composed of several proxies located in different virtual machines that have diversity in their operating systems and are managed by a controller. Proxies can have three modes, namely online, offline, and cleansing. The number of online proxies can be one or more based on the decision of the controller. Depending on a defined exposure time for proxies and a round-robin algorithm, the controller deals with the rotation and changing turn between proxies\citep{Bangalore2009a}. When the exposure time requirement for a proxy is met, it will go through the rejuvenation process (or cleansing process) and will be in cleansing mode. Then, its mode will be altered to offline mode and it will be ready to be chosen by the controller to go online.
\subsection{An Attack Scenario}
We can describe the working principle of the proposed ITS architecture by an attack scenario. Suppose a possible intrusion scenario in which an attacker (an outsider or a malicious insider that has gained access to the SCADA system in smart grid and tries to infect one or more replicas of a critical component (the number of compromised replicas are less than or equal to $f$). Thus he/she would be able to issue control commands. It is also possible that the adversary makes the replica work not properly (e.g., by running a Trojan or changing some system files) which may result in sending inappropriate commands (in case of automatic operation). However, the command must first pass the compromised/faulty replica detector. It is highly probable that the compromised replica(s) being recognized by the inspector (using detection capabilities) or voting (due to the fact that the replicas have different operating systems, all of them may not be infected by the same attack targeted at a special type of vulnerability, and thus the generated responses would be different), so the command will not go further and the infected replicas will undergo reactive recovery. In addition, process manager running in each replica may detect the infection and trigger the process level rejuvenation. Even if the intrusion tolerance mechanisms fail to detect the intrusion, it is possible that the attack's impact is masked through proactive recovery.
\section{Performance Analysis of the Proposed ITS Architecture}
\label{sec: performance analysis}
Security quantification of the proposed ITS architecture is needed for assessing the outcome of the desired performance measures as well as performance comparison with other architectures. To achieve this goal, a state-space model is developed that incorporates an attacker's behavior along with the system's response to an attack or intrusion\citep{Uemura2010,Madan2004167,Nguyen2009,Griffin2005c}. State transition diagrams assist in the evaluation of the transitions impacted by the inter-domain dependencies in the cyber-physical systems. They describe how the attacker's actions cause transitions to failure states\citep{Sridhar2012}. The main advantage of state transition models is the ability to provide a fine-granular system description which includes the dynamic behavior of system\citep{Helvik2008209}. Moreover, these models are tailored to model immense and complex systems such as the smart grid. Markov chain is the basis for diverse state-space techniques in dependability analysis\citep{Zeng2012}. Markov models have often been adopted for software and hardware performance and dependability evaluation following from their capability to capture a variety of dependencies and the simplicity to compute steady-state, transient, and cumulative transient measures. Semi-Markov Process (SMP) is a generalization of both continuous and discrete time Markov chains which allows arbitrary state holding time distribution functions, probably relying on both the current state and on the state to be visited afterwards\citep{Distefano2012}.

Availability and reliability analysis of the smart grid control center networks using Stochastic Petri Nets (a kind of state-space models) has been provided in\citep{Zeng2012}. In this paper, we place focus on the security analysis of our proposed ITS for smart grid control centers with regard to availability and Mean Time To Security Failure (MTTSF) as performance measures using a semi Markov model. The analytical evaluation has been carried out using MATLAB simulator.
\subsection {System Model}
\begin{figure}[!t]
\centering
\includegraphics[width=3.4in]{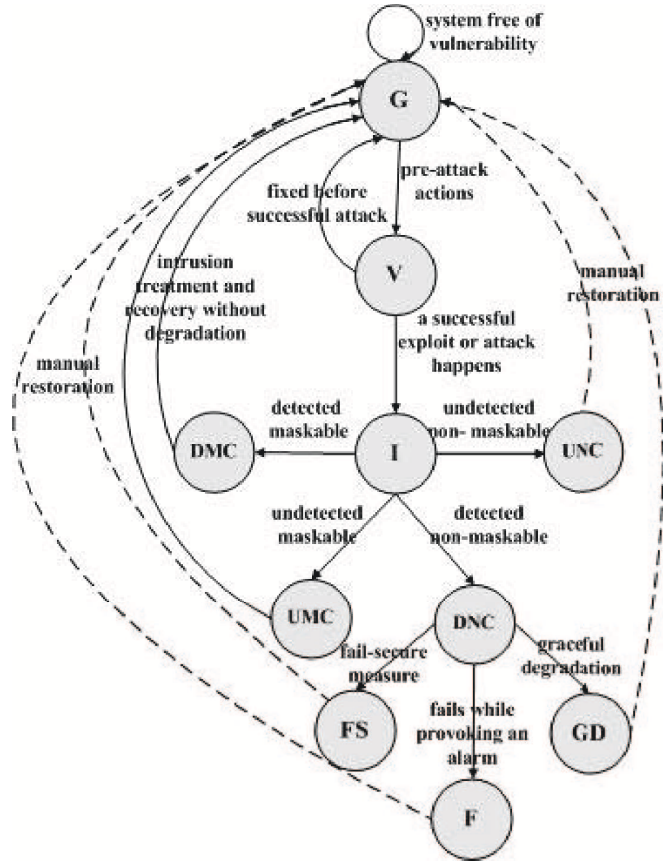}
\caption{State transition diagram for the proposed ITS.}
\label{fig:STD}
\end{figure}

The state transition diagram derived is shown in Figure~\ref{fig:STD} and serves as a generic model for analyzing the behavior of various ITS architectures, including the proposed architecture. It incorporates different security related states of the ITS and their respective interrelationships. Table
\ref{table: state descriptions} presents these security states and their corresponding descriptions.
The system changes from one state to the other during its functional lifespan following from normal usage, abuse, maintenance and corrective measures, failures, and so on. Therefore, the behavior of the system is portrayed as the transitions between the states and each transition corresponds to a specific event. Since the interval between the transition from one state to the other (i.e., state holding time or inter event time) is inclined to be random, its underlying process is defined as a stochastic process \citep{Helvik2008209}. In our system, this process is associated with arbitrary probability distributions, thus, it can be modeled using an SMP.
\begin{table}
\renewcommand{\arraystretch}{1.3}
\caption{Different states of the system and their respective descriptions.}
\label{table: state descriptions}
\centering
\begin{tabular}{c||c}
\hline
\bfseries State & \bfseries Description \\
\hline\hline
G & Good  \\
\hline\hline
V & Vulnerable \\
\hline\hline
I & Intruded \\
\hline\hline
DMC & Detected Masked Compromised \\
\hline\hline
UMC &Undetected Masked Compromised \\
\hline\hline
UNC & Undetected Not masked Compromised \\
\hline\hline
DNC &Detected Not masked Compromised \\
\hline\hline
GD& Graceful Degradation\\
\hline\hline
FS& Fail-secure\\
\hline\hline
F& Failed\\
\hline
\end{tabular}
\end{table}
We can classify the possible transitions in Figure~\ref{fig:STD} according to their starting states as follows:
\begin {enumerate}
\item Transition from the state G: A system free of vulnerabilities is envisioned as being in good state G. During the probing and scanning the system, the identification of vulnerabilities, makes it possible for an adversary to evade or overcome prevention and detection mechanisms and violate the system's security policy. As a result, the system state changes from good state G to the vulnerable state V. Even if the system possesses potential vulnerabilities that may be abused by the malicious intent, it can be regarded as being in the vulnerable state.
\item Transitions from the state V: Discovering a vulnerability (i.e., before an intrusion) and subsequently fixing it brings the system from the state V into the state G. The other possible transition occurs following from the successful exploitation of a vulnerability and will lead to the intruded state I.
\item Transitions from the state I: There are four feasible transitions to other states from the state I. First, if the intrusion tolerance techniques employed in the ITS fail to detect the intrusion and mitigate the impacts of an attack (i.e., mask the attack’s impacts), the system goes to the state UNC with no service guarantee. Second, if the intrusion is detected and the intrusion tolerance techniques succeed in masking the attack' s impact, the state of the system will change from I to DMC. In this state, the intrusion is handled by faulty/compromised replica detector and rejuvenation modules. Third, if the intrusion goes undetected but masked through proactive recovery, a transition to state UMC is made. Subsequently, restoration without any service degradation enables reaching the state G from states UMC or DMC. This is where our state diagram differs from\citep{Madan2004167} in which the ITS architecture (i.e., SITAR) did not possess proactive recovery (corresponding to state UMC in our system). More specifically, the audit module in SITAR carries out periodic diagnosis tests to verify the correct operation of other components and forwards the results to adaptive reconfiguration module to take appropriate actions\citep{Wang2003a} which may include some type of reactive recovery. The last possible transition is to the state DNC when an intrusion is identified but the containment of the damage fails.
\item Transitions from the state DNC: It is possible for an attacker to be able to compromise more than $f$ replicas (e.g., in case of a DoS attack). This may result in complete system failure (transition from DNC to F) or entering states GD or FS. In the state GD, the system is only able to provide essential services which might have different definitions in various systems whereas in the state FS, the system would stop functioning.
\item Transitions from F, GD, FS and UNC: The endpoint of all these transitions will be the state G. The aforementioned transitions would involve manual restoration and corrective maintenance.
\end{enumerate}
\subsection{SMP Analysis}
An SMP can be studied by finding the embedded discrete time Markov chain that requires two sets of parameters \citep{Madan2004167}\citep{Griffin2005c}:
\begin{enumerate}
\item mean sojourn time (i.e., state holding time) for each state
\item the transition probabilities between different states
\end{enumerate}
With respect to Figure~\ref{fig:STD}, the Discrete Time Semi Markov Model (DTSMM) possesses a discrete state space $X_s$= \{G, V, I, UMC, DMC, DNC, UNC, FS, F\} for which $h_i$ indicates the mean sojourn time in state $i \in X_s$ and $p_{ij}$ represents the transition probabilities between states $i$ and $j$ ($i, j \in X_s$).
\subsection {Availability Formulation and Analysis}
Availability and service continuity as the most vital security attribute of the smart grid is required to be analyzed and evaluated for the proposed ITS architecture. Using an SMP model and the steady-state probabilities of its states assists in the steady-state availability analysis of the proposed ITS. We analyze the sensitivity of the availability with respect to two parameters, including the probability of intrusion ($p_{\scalebox{0.5}{I}}$) and the mean time to resist becoming vulnerable to intrusions ($h_{\scalebox{0.5}{G}}$). In addition, a comparison between the proposed ITS and two of the well-known existing ITSs, namely SITAR and SCIT has been made using these parameters.

The steady-state availability $A$ is defined as the probability that the system is in one of normal functioning states. One approach to determine the availability is to pinpoint what the unavailable states (i.e., states FS, F and UNC) are. Thus, the steady-state availability $A$ can be formulated as,
\begin{equation} \label{eq: availability}
A=1-(\pi_{\scalebox{0.5}{UNC}} + \pi_{\scalebox{0.5}{FS}}+ \pi_{\scalebox{0.5}{F}})
\end{equation}
where $\pi_{i}$, $i \in \{UNC, FS, F\}$ denotes the steady-state probability of being in state $i$ for the SMP, that can be computed as,
\begin{equation} \label{eq: steady state availability in state i}
\pi_{i} = \frac{\nu_{i}h_{i}}{\sum \nu_{j}h_{j}}, \quad i,j \in X_{s}
\end{equation}
where $h_i$ indicates the mean state holding time in state $i$ and $v_i$ denotes the embedded Discrete Time Markov chain (DTMC) steady-state probability in state $i$. We can derive $v_i$s from the following two equations,
\begin{equation} \label{eq: v.p}
\nu = \nu \cdot P
\end{equation}
\begin{equation} \label{eq: sum of vi}
\sum_{i} \nu_{i} = 1, \quad i \in X_{s}
\end{equation}
where the $P$ is the transition probability matrix of the corresponding DTMC for the proposed ITS,

\begin{center}
\scalebox{0.9}{
 {$P$} = $\bordermatrix{
        &  G &  V &  I &  DMC &  UNC &  UMC &  DNC &  FS &  GD &  F \cr
      G & 0 & 1 & 0 & 0 & 0 & 0 & 0 & 0 & 0 & 0  \cr
      V & 1-p_{\scalebox{0.5}{I}} & 0 & p_{\scalebox{0.5}{I}} & 0 & 0 & 0 & 0 & 0 & 0 &
      0  \cr
      I & 0 & 0 & 0 & p_{\scalebox{0.5}{DM}} & p_{\scalebox{0.5}{UN}} &     
      p_{\scalebox{0.5}{UM}} & p_{\scalebox{0.5}{DN}} & 0 & 0 & 0  \cr
      DMC & 1 & 0 & 0 & 0 & 0 & 0 & 0 & 0 & 0 & 0  \cr
      UNC& 1 & 0 & 0 & 0 & 0 & 0 & 0 & 0 & 0 & 0  \cr
      UMC & 1 & 0 & 0 & 0 & 0 & 0 & 0 & 0 & 0 & 0  \cr
      DNC & 0 & 0 & 0 & 0 & 0 & 0 & 0 & p_{\scalebox{0.5}{FS}} & p_{\scalebox{0.5}{GD}} & p_{\scalebox{0.5}{F}}  \cr
      FS & 1 & 0 & 0 & 0 & 0 & 0 & 0 & 0 & 0 & 0   \cr
      GD & 1 & 0 & 0 & 0 & 0 & 0 & 0 & 0 & 0 & 0  \cr
      F & 1 & 0 & 0 & 0 & 0 & 0 & 0 & 0 & 0 & 0  \cr
      }$
}
\end{center}

In this research, the mean state holding times $h_i$ for all the states of DTMC have been assumed to have the same values as\citep{Madan2004167} except for the state UMC which is a new state (corresponding to proactive recovery) for our proposed ITS.

Finally, by using (1)-(4), the steady-state availability ($A_{\scalebox{0.5}P}$) of our proposed ITS is computed as,
\begin{equation}  \label{eq: proposed ITS availability}
A_{\scalebox{0.5}p} =1-\frac{h_{\scalebox{0.5}{UNC}}p_{\scalebox{0.5}{I}}p_{\scalebox{0.5}{UN}}+h_{\scalebox{0.5}{F}}p_{\scalebox{0.5}{I}}p_{\scalebox{0.5}{DN}}p_{\scalebox{0.5}{F}} +h_{\scalebox{0.5}{FS}}p_{\scalebox{0.5}{I}}p_{\scalebox{0.5}{DN}}p_{\scalebox{0.5}{FS}}}{h_{\scalebox{0.5}{G}}+h_{\scalebox{0.5}{V}}+p_{\scalebox{0.5}{I}}(h_{\scalebox{0.5}{I}}+h_{\scalebox{0.5}{DMC}}p_{\scalebox{0.5}{DM}}+h_{\scalebox{0.5}{UNC}}p_{\scalebox{0.5}{UN}}+h_{\scalebox{0.5}{UMC}}p_{\scalebox{0.5}{UM}} 
+h_{\scalebox{0.5}{DNC}}p_{\scalebox{0.5}{DN}}+h_{\scalebox{0.5}{GD}}p_{\scalebox{0.5}{DN}}p_{\scalebox{0.5}{GD}}
+h_{\scalebox{0.5}{FS}} p_{\scalebox{0.5}{DN}}p_{\scalebox{0.5}{FS}}+ h_{\scalebox{0.5}{F}}p_{\scalebox{0.5}{DN}}p_{\scalebox{0.5}{F}})} 
\end{equation}
In a similar manner, the steady-state availability for SITAR ($A_{\scalebox{0.5}{SITAR}}$) and SCIT ($A_{\scalebox{0.5}{SCIT}}$) are derived as,
\begin{equation}\label{eq: SITAR availability}
A_{\scalebox{0.5}{SITAR}} = 1-
 \frac{h_{\scalebox{0.5}{UNC}}p_{\scalebox{0.5}{I}}p_{\scalebox{0.5}{UN}+h_{F}}p_{\scalebox{0.5}{I}}p_{\scalebox{0.5}{DN}}p_{\scalebox{0.5}{F}} +h_{\scalebox{0.5}{FS}}p_{\scalebox{0.5}{I}}p_{\scalebox{0.5}{DN}}p_{\scalebox{0.5}{FS}}}{h_{\scalebox{0.5}{G}}+h_{\scalebox{0.5}{V}}+p_{\scalebox{0.5}{I}}(h_{\scalebox{0.5}{I}}+h_{\scalebox{0.5}{DMC}}p_{\scalebox{0.5}{DM}}+h_{\scalebox{0.5}{UNC}}p_{\scalebox{0.5}{UN}}+
h_{\scalebox{0.5}{DNC}}p_{\scalebox{0.5}{DN}}
+ h_{\scalebox{0.5}{GD}}p_{\scalebox{0.5}{DN}}p_{\scalebox{0.5}{GD}}
+ h_{\scalebox{0.5}{FS}}p_{\scalebox{0.5}{DN}}p_{\scalebox{0.5}{FS}} +h_{\scalebox{0.5}{F}}
p_{\scalebox{0.5}{DN}}p_{\scalebox{0.5}{F}})}
\end{equation}
\begin{equation} \label{eq: SCIT availability}
A_{\scalebox{0.5}{SCIT}} = 1-\frac{h_{\scalebox{0.5}{F}}p_{\scalebox{0.5}{I}}p_{\scalebox{0.5}{F}}}{h_{\scalebox{0.5}{G}}+h_{\scalebox{0.5}{V}}+p_{\scalebox{0.5}{I}}(h_{\scalebox{0.5}{I}}+h_{\scalebox{0.5}{UMC}}p_{\scalebox{0.5}{UM}}+h_{\scalebox{0.5}{F}}p_{\scalebox{0.5}{F}})}
\end{equation}

It should be pointed out that some of the transition probabilities may have different values or even may not be applicable for all three ITSs following from the fact that the three ITSs do not possess the same state space (DTSMM' s state space for SITAR does not include state UMC whereas SCIT does not contain the states DMC, DNC, UNC, GD and FS).

\begin{figure}[!t]
\centering
\includegraphics[width=3.4in]{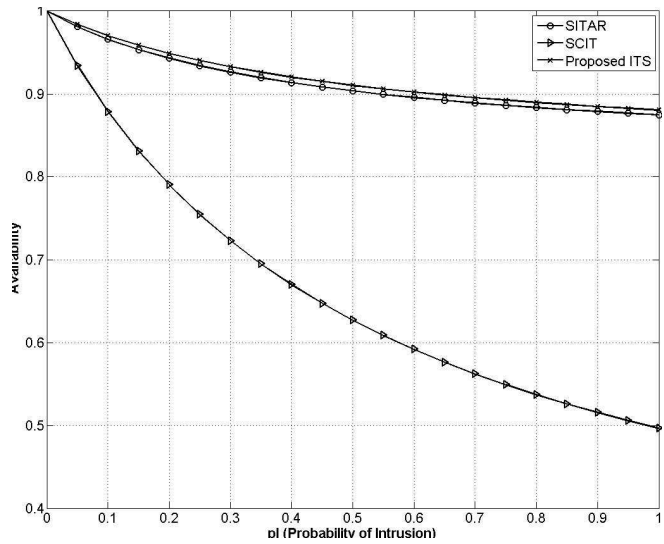}
\caption{Availability vs $p_{I}$.}
\label{fig:AvailbilitypI}
\end{figure}

\begin{figure}[!t]
\centering
\includegraphics[width=3.4in]{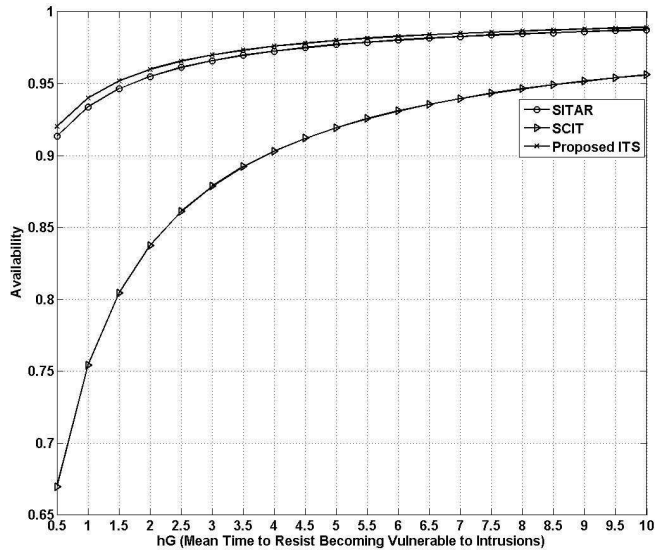}
\caption{Availability vs $h_{G}$.} 
\label{fig:AvailbilityhG}
\end{figure}

Figure~\ref{fig:AvailbilitypI} and Figure~\ref{fig:AvailbilityhG} illustrate the availability performance of the proposed ITS, SITAR and SCIT with regard to $p_{\scalebox{0.5}{I}}$ and $h_{\scalebox{0.5}{G}}$ respectively. The steady-state availability is a decreasing function of $p_{\scalebox{0.5}{I}}$ for all three ITSs (Figure~\ref{fig:AvailbilitypI}). The availability for SCIT falls sharply when the probability of intrusion increases compared to the other two ITSs. This is due to the fact that SCIT lacks detection capabilities and only uses periodic rejuvenation. Considering Figure~\ref{fig:AvailbilitypI}, availability performance of the proposed ITS shows 0.6\% and 36\% improvement compared to SITAR and SCIT respectively (using the same values for parameters). Figure~\ref{fig:AvailbilityhG} shows the positive impact of increasing the time that the system is in the good state on the availability (i.e., the availability increases as the $h_{\scalebox{0.5}{G}}$ rises). For larger amount of $h_{\scalebox{0.5}{G}}$, there is a slight difference in availability performance of the three ITS. In this figure, availability performance of the proposed ITS presents 0.3\% and 9\% improvement compared to SITAR and SCIT respectively. This is mostly due the use of the hybrid and fine-grained recovery approach in the proposed ITS that contributes to improve the system's availability.

\subsubsection{SLA as another possible performance criterion}
Service Level Agreement (SLA) can be considered as another performance measure in critical infrastructures such as smart grid. The SLAs (service level agreements) are pre-defined agreements on some of the QoS parameters, including response time, delay, data rate, and so on. Considering SLAs based on response time can be a true assumption since all SLAs are expected to improve the observed response time. As stated in\citep{VasireddyR.andTrivedi2006}, having five nines availability does not suggest the guarantee of the system' s SLA. This means that even if a system is available most of the time, it may not meet the SLA requirements. This can be applied to the smart grid control centers in which satisfying the delay requirements is of utmost importance. Therefore, similar to the definition of the steady-state availability, the steady-state service level agreeability has been defined in\citep{VasireddyR.andTrivedi2006}. They divided the sates of the system according to a threshold response time. The viable sets were highly SLA satisfying, SLA satisfying, SLA violating and highly SLA violating. Using the steady-state service level agreeability concept, we can group our state diagram states into the four aforementioned clusters. It is evident that in states G and V the system can satisfy the SLA completely while in the intruded state I and states DMC and UMC in which intrusion tolerance mask measures are taken, SLA satisfaction may not be the same as highly SLA satisfying class. In state DNC, the intrusion has been detected but cannot be masked and also in sate GD in which the system only provides the essential services, we expect degradatrion of service, thus the SLA is violated. Finally, state FS requires the system to stop functioning and states UNC and F that will result in the security failure of the system fall into the highly SLA violating group. To obtain the steady-state service level agreeability, we can compute the steady-state probabilities of the states using the SMP model and get a summation of these probabilities for the states included in each cluster.
\subsection{MTTSF Formulation and Analysis}
Analogous to the Mean Time To Failure (MTTF) as a quantitative reliability measure, MTTSF is a measure for quantifying the security of intrusion tolerant systems\citep{Madan2004167}. MTTSF is defined as the mean elapsed time for the system to reach one of the security-compromised states (also called absorbing states), provided that the system begins in state G. Using a similar approach to availability analysis, we analyze the MTTSF with regard to $p_{\scalebox{0.5}{I}}$ and $h_{\scalebox{0.5}{G}}$ parameters. We take advantage of an SMP with absorbing and transient states. In the state transition diagram shown in Figure~\ref{fig:STD}, the set of states $X_a$ = \{UNC, GD, FS, F\} are considered as the absorbing states (i.e., the probability of moving out of these states is zero). These states indicate the security compromised states. The rest of the states are called transient states and denoted by $X_t$ = \{G, V, I, UMC, DMC, DNC\}. The transition probability Matrix $M$ exhibits the transition probabilities between the transient states (i.e., $Q$) and the states originating from transient states to absorbing states (i.e., $C$) in an organized form.

\begin{center}
 {$M$} =$
\begin{pmatrix}
Q & | & C \cr
- - & | & - - \cr
0 & | & I \cr
\end{pmatrix}
$
\end{center}

Matrixes $Q$ and $C$ are as follows:

\begin{center}
{$Q$} = $\bordermatrix{
        &  G &  V &  I &  DMC  &  UMC &  DNC  \cr
      G & 0 & 1 & 0 & 0 & 0 & 0   \cr
      V & 1-p_{\scalebox{0.5}{I}} & 0 & p_{\scalebox{0.5}{I}} & 0 & 0 & 0   \cr
      I & 0 & 0 & 0 & p_{\scalebox{0.5}{DM}} & p_{\scalebox{0.5}{UM}} &    p_{\scalebox{0.5}{DN}}   \cr
      DMC & 1 & 0 & 0  & 0 & 0 & 0   \cr
      UMC & 1 & 0 & 0  & 0 & 0 & 0   \cr
      DNC & 0 & 0 & 0  & 0 & 0 & 0  \cr
                }$
\end{center}

\begin{center}
{ $C$} = $\bordermatrix{
        & UNC & FS &  GD &  F \cr
      G & 0 & 0 & 0 & 0  \cr
      V & 0 & 0 & 0 & 0   \cr
      I &  p_{\scalebox{0.5}{UN}} & 0 & 0 & 0  \cr
      DMC & 0 & 0 & 0 & 0   \cr
      UMC & 0 & 0 & 0 & 0   \cr
      DNC & 0 & p_{\scalebox{0.5}{FS}} & p_{\scalebox{0.5}{GD}} & p_{\scalebox{0.5}{F}}  \cr
            }$ 
\end{center}

As stated in\citep{Madan2004167}, we can find the MTTSF by the following formula,
\begin{equation}  \label{eq: 8}
MTTSF=  \sum_{i \in X_t} V_{i}h_{i}
\end{equation}
where $V_{i}$ indicates the average number of times the transient state $i$ has been visited before the DTMC arrives at one of the absorbing states and $h_{i}$ indicates the mean state holding time in state $i$.

Let $q_i$ be the probability of start in state $i$ (here, it is assumed that the DTMC starts in state G) and  $q_{ji}$  be the transition probability from the transient state $j$ to the transient state $i$. So, the $V_i$s can be computed through solving the system of equations,
\begin{equation}  \label{eq: system of equations}
V_i = q_i + \sum_{j} V_j q_{ji}, \quad i, j \in X_t
\end{equation}
Finally, we use (8) to calculate the MTTSF for the proposed ITS as,
\begin{equation}  \label{eq: proposed ITS MTTSF}
M_{\scalebox{0.5}{P}}=
\frac{h_{\scalebox{0.5}{G}}p_{\scalebox{0.5}{I}}^{-1}+h_{\scalebox{0.5}{V}}p_{\scalebox{0.5}{I}}^{-1}+h_{\scalebox{0.5}{I}}+h_{\scalebox{0.5}{DMC}}p_{\scalebox{0.5}{DM}}+h_{\scalebox{0.5}{UMC}}p_{\scalebox{0.5}{UM}}+h_{\scalebox{0.5}{DNC}}p_{\scalebox{0.5}{DN}}}{1-p_{\scalebox{0.5}{DM}}-p_{\scalebox{0.5}{UM}}}
\end{equation}\\
Using the same approach, we can find the formula for SITAR\citep{Madan2004167} and SCIT,
\begin{equation}  \label{eq: SITAR MTTSF}
M_{\scalebox{0.5}{SITAR}}=\frac{h_{\scalebox{0.5}{G}}p_{\scalebox{0.5}{I}}^{-1}+h_{\scalebox{0.5}{V}}p_{\scalebox{0.5}{I}}^{-1}+h_{\scalebox{0.5}{I}}+
 h_{\scalebox{0.5}{DMC}}p_{\scalebox{0.5}{DM}}+h_{\scalebox{0.5}{DNC}}p_{\scalebox{0.5}{DN}}}{1-p_{\scalebox{0.5}{DM}}}
\end{equation}
\begin{equation}  \label{eq: SCIT MTTSF}
M_{\scalebox{0.5}{SCIT}}  =\frac{h_{\scalebox{0.5}{G}}p_{\scalebox{0.5}{I}}^{-1}+h_{\scalebox{0.5}{V}}p_{\scalebox{0.5}{I}}^{-1}+h_{\scalebox{0.5}{I}}+h_{\scalebox{0.5}{UMC}}p_{\scalebox{0.5}{UM}}}{1-p_{\scalebox{0.5}{UM}}}
\end{equation}

\begin{figure}[!t]
\centering
\includegraphics[width=3.4in]{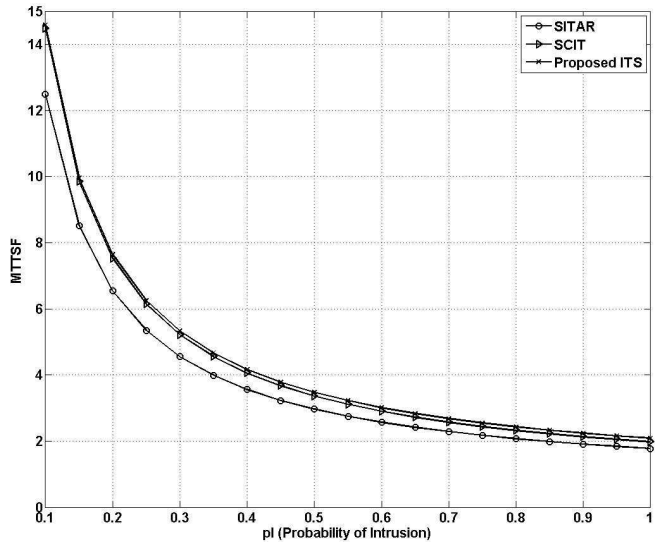}
\caption{MTTSF vs $p_{I}$.}
\label{fig:MTTSFpI}
\end{figure}

\begin{figure}[!t]
\centering
\includegraphics[width=3.4in]{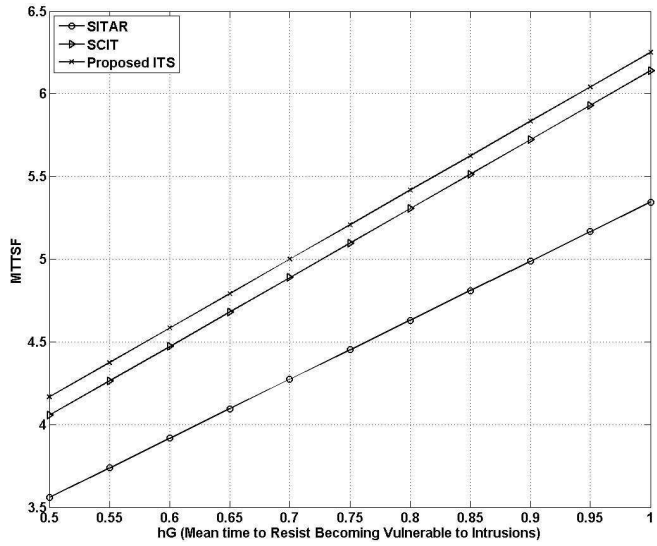}
\caption{MTTSF vs $h_{G}$.}
\label{fig:MTTSFhG}
\end{figure}

As Figure.~\ref{fig:MTTSFpI} illustrates, MTTSF has a reciprocal relationship with the probability of intrusion, i.e., it decreases as the probability of intrusion rises. The proposed ITS architecture shows improved MTTSF with regard to $p_{\scalebox{0.5}{I}}$ since it has more possible security features (e.g., proactive and reactive recovery) and thus more system states (corresponding to tolerance measures) when dealing with intrusions. It demonstrates advance in MTTSF performance compared to other two ITSs (17\% compared to SITAR and 2\% compared to SCIT). Figure~\ref{fig:MTTSFhG} depicts the impact of increasing $h_{\scalebox{0.5}{G}}$ on MTTSF. It is obvious that MTTSF ascends when the system spends more time in state G. In this graph with assigned values to transition probabilities and state holding times, the proactive rejuvenation in SCIT seems to have more effects on the MTTSF when increasing the $h_{\scalebox{0.5}{G}}$ in comparison with the reactive rejuvenation in SITAR. The acquired results show that the stability of our proposed ITS is better than the others. The improvement in MTTSF performance is 16\% and 0.8\% compared to SITAR and SCIT respectively. The acquired results for MTTSF also prove the security enhancement of the proposed architecture (mostly thanks to the combined rejuvenation approach) compared with the other two systems.
\subsection{Discussion and analysis}
As we know, self-healing capability is one of the distinctive features of the smart grid. From the acquired results in the previous sections, we can infer that the mask measures (reflected in the states DMC and UMC in the provided state transition diagram) and in particular, the self-healing capabilities (automatic rejuvenation) of the proposed ITS influences its performance to a considerable extent. From the Figure~\ref{fig:STD}, we have,
\begin{equation}  \label{eq: sum of probabilities}
p_{\scalebox{0.5}{DM}}+p_{\scalebox{0.5}{UM}}+p_{\scalebox{0.5}{UN}}+p_{\scalebox{0.5}{DN}}=1
\end{equation}
By considering $p_{\scalebox{0.5}{M}}$ (i.e., probability of masking an intrusion) as the sum of $p_{\scalebox{0.5}{DM}}$ and $p_{\scalebox{0.5}{UM}}$ as well as $p_{\scalebox{0.5}{N}}$ (i.e., probability of the inability to mask an intrusion) as the sum of $p_{\scalebox{0.5}{DN}}$ and $p_{\scalebox{0.5}{UN}}$, we will have the following equation,
\begin{equation}  \label{eq: sum of mask and non-mask probs}
p_{\scalebox{0.5}{M}}+p_{\scalebox{0.5}{N}}=1
\end{equation}
A perfect and ideal ITS is expected to have the $p_{\scalebox{0.5}{M}}$ equal to one. Therefore, we should attempt to enhance the ITS masking capabilities in order to have a more robust and secure ITS architecture. In the proposed ITS architecture, we made an effort to increase the $p_{\scalebox{0.5}{M}}$ compared with the other two ITSs (i.e., SITAR and SCIT).
\section{Conclusion and future work} \label{sec: conclusion}
This paper has provided an in-depth research on the significance of using intrusion tolerance as a promising security approach to improve the security of smart grid control centers. An ITS architecture was proposed to be adopted in control centers' critical components, particularly SCADA/EMS. Using different intrusion tolerance techniques such as replication, diversity, proactive and fine-grained reactive recovery made the proposed ITS outperform two of well-known architectures, namely SITAR and SCIT. SITAR only possesses reactive recovery and SCIT leverages the periodic rejuvenation. In addition, in our proposed ITS, acceptance testing is only carried on the outgoing data in contrast with SITAR which applies acceptance testing to both incoming and outgoing data. So, the response time of the proposed ITS architecture is expected to decrease. Thus enhancing the security of the proposed ITS. The availability and MTTSF performance measures were analyzed via a Discrete Time Semi Markov Model (that can be used as a general model for assessing the security attributes of any ITS) and compared with other ITSs. In future, we will make an attempt to simulate the proposed  architecture and evaluate its performance with respect to other performance metrics such as delay which is a critical feature for smart gird control centers.

\end{document}